\documentclass{article}
\usepackage[english]{babel}
\usepackage{blindtext}
\usepackage[utf8]{inputenc}
\usepackage{fancyhdr}
\usepackage{booktabs}
\usepackage{tabu}
\usepackage[T1]{fontenc}
\usepackage{titling}

\usepackage{amsmath}
\usepackage{graphicx}
\usepackage[colorinlistoftodos]{todonotes}
\usepackage[colorlinks=true, allcolors=blue]{hyperref}
\usepackage{authblk}

\usepackage{lipsum}
\usepackage{titlesec}
\usepackage[a4paper, total={7.0in, 10.5in}]{geometry}

\title{Pushing compute and AI onto detector silicon\\
AI@DOE Roundtable Whitepaper 2021}

\author[1]{Antonino Miceli \thanks{amiceli@anl.gov}}
\author[2]{Kazutomo Yoshii \thanks{kazutomo@mcs.anl.gov}}
\author[3]{Ian T. Foster \thanks{foster@anl.gov}}
\affil[1]{X-ray Science Division,\
Argonne National Laboratory, Lemont, IL USA}
\affil[2]{Mathematics \& Computer Science Division,\
Argonne National Laboratory, Lemont, IL USA}
\affil[3]{Data Science \& Learning Division,\
Argonne National Laboratory, Lemont, IL USA}
\date{12/08/2021}                     

\begin{document}
\maketitle
\vspace{-30pt}

\section{Challenge}
Pixel detectors are at the heart of advances in imaging. Astronomy has been revolutionized by the use of high-throughput detectors for surveying supernovas, shedding light on expansion rates of the universe, and exoplanet searches using the wobble of faint stars. Biochemistry has been transformed first by developing megapixel detectors for X-ray crystallography, and more recently, by using high frame rate detectors to correct for drift and enable atomic resolution structure determination in cryo-electron microscopy. Materials scientists are gaining new insights into subtle electron bonding arrangements using high dynamic range electron detectors as part of phasing coherent diffraction patterns.  Beyond scientific advances, the photography industry has been revolutionized by the transition to electronic image detectors. 

However, in order to take full advantage of the DOE’s billion-dollar investments into the next-generation research infrastructure (e.g., exascale, light sources, colliders), advances are required not only in detector technology but also in computing and specifically AI. Let us consider an example from X-ray science. Nanoscale X-ray imaging is a crucial tool to enable a wide range of scientific explorations from materials science and biology to mechanical and civil engineering. The next-generation light sources will increase the X-ray beam brightness and coherent flux by 100 to 1,000 times. In order to image larger samples, the continuous frame rate of pixel array detectors must be increased, approaching 1 MHz, which requires several Tbps (aggregated) to transfer pixel data out to a data acquisition system. Using 65-nm CMOS technology, an optimistic raw data rate off such a chip is about 100-200 Gbps. However, a continuous 1 MHz detector with only $256 \times 256$  pixels at 16-bit resolution, for example, will require 1,000 Gbps (i.e., 1 Tbps) bandwidth off the chip!  It is impractical to have multiple high-speed transceivers running in parallel to provide such bandwidth and represents the first data bottleneck. New approaches are necessary to reduce the data size by performing data compression or AI-based feature extraction directly inside a detector silicon chip in a streaming manner before sending it off-chip.

The dividing line between edge and exascale computing is often ambiguous. However for scientific imaging detectors, the ultimate location to incorporate edge compute is directly on the front-end detector silicon.  Many of these scientific detectors are constructed from application-specific integrated circuits (ASICs). Until recently, pixel detector ASICs have been used mainly for analog signal processing of the charge from the sensor layer and the transmission of raw pixel data off the detector ASIC. However, with the availability of more advanced ASIC technology nodes for scientific application, more digital functionality from the computing domains (e.g., compression and feature extraction) can be  integrated directly into the detector ASIC (e.g., on the chip's periphery or edge) to overcome off-chip bandwidth limitation and ultimately increase data velocity. This increase in data velocity will not only enable observation phenomena at faster timescales, but also enable smarter experiments to more quickly focus on important regions of interest and detect rare events. We should stress that an increase in data velocity can only be achieved by embedding AI into the detector silicon itself. External AI accelerators (e.g., GPUs, TPUs, etc) can only help with downstream bottlenecks and will not fundamentally increase data velocity.

\section{Opportunity}
The scientific pixel detectors are typically built using ASICs. In general, integrated circuits are at the core of smartphones and cameras, the internet of things, computers, cloud data centers, and a new wave of artificial intelligence processors. What has enabled the proliferation of integrated circuits beyond traditional computing units (i.e., CPUs) is the “pure-play” semiconductor foundry business model, where even small-scale companies can develop designs that are then fabricated using billion-dollar semiconductor fabrication facilities. This resulted in a new opportunity for fabless semiconductor companies to emerge which focus only on integrated circuit design. The academic research community is fortunate to have access to these foundries through third-party brokers who facilitate cost-effective prototyping of small designs and full-scale fabrications at  advanced technology nodes. With the availability of these advanced technology nodes, the number of digital logic resources available increases. The use of digital circuitry also opens up the possibility of incorporating  digital data processing (e.g., compression, machine learning) directly on the detector ASIC.

The current generation of electron and X-ray pixel detectors and those on the horizon work at kHz frame rates in continuous mode. Pixel detectors also exist that capture images at $\sim$~10 MHz frame rates in burst mode. For example, the EU-XFEL’s AGIPD and LPD detectors and Cornell University’s Keck-PAD use on-chip analog storage to capture bursts of 8 to 384 images at MHz rates but  are subsequently read out slowly.  These detector projects demonstrate that the current generation of analog front ends can handle $\sim$~10 MHz frame rates. The common bottleneck of these detectors is the limited data bandwidth off the detector ASIC resulting from the analog design flow. Digital technology can increase the off-chip bandwidth if we follow a more modern mixed-signal ASIC design flow methodology using sub-100 nm CMOS processes. 

While recent exploratory work to develop compute on detector ASICs is promising~\cite{Hammer_2021,Yoshii_codesign,Strempfer:2021uo}, much more is required. A co-design process that brings together ASIC designers, computer and application scientists is critical to make significant progress. In addition, a DOE-wide effort is needed to merge compute with detector ASICs. However, a big hurdle to cross-lab efforts are the significant legal issues with NDAs from chip foundries as well as the very high entry to barrier of commercial electronic design automation (EDA) tools since academic pricing is not offered to DOE labs. Ideally, one would like to have affordable EDA tools and NDAs where all the DOE labs can share circuit designs which each other, just like computer scientists have done with open-source software. A substantial coordinated effort is required to address these issues. Technological innovation should be centered around ultrafast inference architecture designs and software technologies to assist hardware development for detectors. This research effort on ultrafast inference architecture could also benefit future high-performance computation as workloads for processing units (e.g., FPU) shift from conventional numerical computing to AI workloads.

\section{Timeliness or maturity}
We believe that now is the perfect time to invest in the development of a new generation of AI-accelerated detectors. Billions of dollars are being spent on next-generation light source and collider facilities in the US (ALS-U, APS-U, CHESS-U, LCLS-II, EIC) and around the world (MAX-IV, ESRF-U, PETRA-IV, Beijing Light Source, etc.) as well as at leadership computing facilities. While the brightness increases with these new  accelerator upgrades will be significant, these new sources will not reach their full potential if limited to using detectors of the type available today, or those presently seen on the horizon. In addition, application scientists have increasingly turned to AI to analyze data~\cite{sanaullah2018real,Abeykoon2019,Zhengchun2019,Cherukara:2020il,liu2020braggnn}. AI-accelerated workflows have been shown not only to be fast enough to keep up with experiments, but also to overcome experimental restrictions of conventional methods. Now is the time to merge compute, AI and detectors!


\bibliographystyle{ieeetr}
\bibliography{refs}

\end{document}